\documentclass[prb,twocolumn,showpacs,superscriptaddress]{revtex4}
\usepackage{graphicx}
\usepackage{color}

\begin{document}

\title{Doping and bond length contributions to Mn K-edge shift in
La$_{1-x}$Sr$_x$MnO$_{3}$ ($x$=0-0.7) and their correlation with
electrical transport properties}

\author{S. K. Pandey}
\altaffiliation[Present address: ]{Department of Condensed Matter
Physics and Material Sciences, Tata Institute of Fundamental
Research, Homi Bhabha Road, Kolaba, Mumbai-400 005, India.}
\email{sk_iuc@rediffmail.com}

\author{R. Bindu}
\altaffiliation[Present address: ]{Material Science Division,
Indira Gandhi Center for Atomic Research, Kalpakkam, India.}
\affiliation{UGC-DAE Consortium for Scientific Research,
University Campus, Khandwa Road, Indore 452 017, India}

\author{Ashwani Kumar}
\altaffiliation[Present address: ]{Department of Physics,
Institute of Science and Laboratory Education, IPS Academy, Indore
452 012, India.} \affiliation{School of Physics, Devi Ahilya
University, Khandwa Road, Indore 452 017, India}

\author{S. Khalid}
\affiliation{National Synchrotron Light Source, Brookhaven
National Laboratory, Upton, NY - 11973, USA}

\author{A. V. Pimpale}
\email{avp@csr.ernet.in} \affiliation{UGC-DAE Consortium for
Scientific Research, University Campus, Khandwa Road, Indore 452
017, India}

\begin{abstract}

The room temperature experimental Mn K-edge x-ray absorption
spectra of La$_{1-x}$Sr$_x$MnO$_{3}$, $x$ = 0 - 0.7 are compared
with the band structure calculations using spin polarized density
functional theory. It is explicitly shown that the observed shift
in the energy of Mn K-edge on substitution of divalent Sr on
trivalent La sites corresponds to the shift in the center of
gravity of the unoccupied Mn 4$p$-band contributing to the Mn K-
absorption edge region. This correspondence is then used to
separate the doping and size contributions to the edge shift due
to variation in number of electrons in valence band and Mn-O bond
lengths, respectively, when Sr is doped into LaMnO$_3$.  Such
separation is helpful to find the localization behaviour of charge
carriers and to understand the observed transport properties of
these compounds.

\end{abstract}


\maketitle

\section{Introduction}

In recent years, hole-doped manganese oxides have been the subject
of intense research because of their various interesting
properties: colossal magnetoresistance (CMR), charge ordering,
orbital ordering, phase separation, {\it etc.}
\cite{imada,coey,dagotto}.  The compounds under present study
La$_{1-x}$Sr$_x$MnO$_3$, $x$ = 0 - 0.7 also show these interesting
properties for specific range of $x$ \cite{hem}. In this doping
range the resistivity behaves in a non-monotonic fashion, first
decreasing as $x$ increases up to 0.4 and then increasing as $x$
further increases for a wide temperature range, see figure
\ref{expt}c. This behaviour is not very well understood. Although
double exchange mechanism is commonly employed, it is known to be
inadequate \cite{millis}. Qualitatively, by considering
Jahn-Teller splitting of $e_g$ band of the Mn and invoking strong
Hund coupling it is seen that Sr doping corresponds to adding
holes in the system. Thus as $x$ increases the resistivity should
decrease for low $x$. Naively one would expect the resistivity to
have a minimum at $x$ = 0.5; however, the detailed interactions
involved in conduction phenomenon could alter this situation.
Besides adding holes to the system, Sr doping would also change
the lattice parameters due to its different ionic radius as
compared to La. The changed overlap integrals would modify the
local electronic densities and localization of electrons and holes
at different sites in the lattice, thus affecting the transport
properties.  Coexistence of localized Jahn-Teller polaronic and
broad band $e_g$ states may also play important role in these
compounds \cite{tvr}.

X-ray absorption (XA) spectra provide invaluable information about
the electronic states of absorbing atom and arrangement of
surrounding atoms.  An XA edge represents the transition of
electron from a core level to the low lying unoccupied states of
appropriate symmetry.  The edge position of an absorption spectrum
depends on electronic charge distribution around the absorbing
atom \cite{azar}. A shift in the absorption edge energy is thus a
manifestation of the change in the effective charge on the
absorbing atom.  The effective charge can be varied by changing
the number of electrons through appropriate doping and also by
changing the interatomic distance. Therefore, in reference to the
system under present study, an energy shift in Mn K-edge position
on replacing some of trivalent La with divalent Sr will have two
contributions: one from the changed number of valence electrons
and other from changes in the Mn-O bond length.  Further, a change
in electronic charge distribution is directly related with
localization behaviour of charge carriers and hence electrical
transport properties. Therefore, a knowledge of changes in the
aforementioned two contributions to the edge shift with Sr
concentration will be helpful in understanding the transport
properties of manganites. There are many works in the literature
reporting XA spectra at Mn K-edge in manganites
\cite{croft,subias,garcia,ignatov,booth,qian}. Almost all this
work is concentrated on La$_{1-x}$Ca$_x$MnO$_3$ and the change in
the edge shift is correlated with the change in the electronic
state of Mn ion on Ca doping.  Only a few of these studies
\cite{ignatov,booth,qian} bring out the importance of lattice and
band structure effects on the absorption edge, estimation of
different contributions to the edge shift and their relation with
various physical properties of manganites still remains a
challenging task.

In this contribution we show using experimental XA spectra of
La$_{1-x}$Sr$_x$MnO$_3$ and electronic structure calculations that
shift in Mn K-edge position on Sr doping in LaMnO$_3$ corresponds
to the changes in the centre of gravity (CG) of the unoccupied Mn
4$p$-band contributing to the Mn K-edge region.  This
correspondence is then used to separate the two contributions to
the edge shift: one from changed number of valence electrons and
the other from changes in Mn-O bond length, which we denote as
doping and size contributions, respectively.  It is observed that
doping contribution to the edge shift with varying Sr
concentration follows the observed electrical transport behaviour,
which is directly related with localization of the charge
carriers, thus emphasizing the importance of hole localization in
governing the transport behaviour of these series of compounds.

\section{Experimental and computational details}

Polycrystalline samples of La$_{1-x}$Sr$_x$MnO$_{3}$, $x$ = 0.0 -
0.7 were prepared by solid state route. The oxygen content in all
the samples was measured by iodometric titration.  While the
samples with $x$ = 0.1 - 0.3 were found to be off-stoichiometric,
the samples with $x$ = 0.4 - 0.7 were stoichiometric within the
experimental accuracy. XA spectra were recorded at room
temperature (300 K) on beamline X-18 B using a Si(111) channel cut
monochromator at the National Synchrotron Light Source, Brookhaven
National Laboratory. The details of sample preparation and
experiments are given elsewhere~\cite{binduxrd,binduexafs}. As
shown in our earlier EXAFS work on La$_{1-x}$Sr$_x$MnO$_3$
\cite{binduexafs}, MnO$_6$ octahedron is distorted throughout the
series and four Mn-O bonds have same bond length and other two
have different bond lengths. Therefore, for the purpose of band
structure calculations, the local structure was represented by a
tetragonal lattice in the whole range, $x$ = 0 to 0.7, with
interchange of lattice parameters $a$ and $c$ at $x$ = 0.4 to
account for the change in local structure around $x$ = 0.4.
Lattice parameters taken for obtaining the muffin-tin radii were
two times of these Mn-O bond lengths. Spin polarized electronic
structure calculations were carried out using LMTART 6.61
\cite{savrasov}. For calculating charge density, full-potential
LMTO method working in plane wave representation was used.  The
charge density and effective potential were expanded in spherical
harmonics up to $l$ = 6 inside the sphere and in a Fourier series
in the interstitial region. The exchange correlation functional of
the density functional theory was taken after Vosko, Wilk, and
Nussair  and the generalized-gradient approximation scheme of
Perdew {\it et al.} \cite{perdew} was also invoked. (6, 6, 6)
divisions of the Brillouin zone along three directions for the
tetrahedron integration were used to calculate the density of
states. Self-consistency was achieved by demanding the convergence
of the total energy to be smaller than 10$^{-4}$ Ry/cell.

\section{Results and discussion}

\begin{figure}
\includegraphics[width=8.5cm]{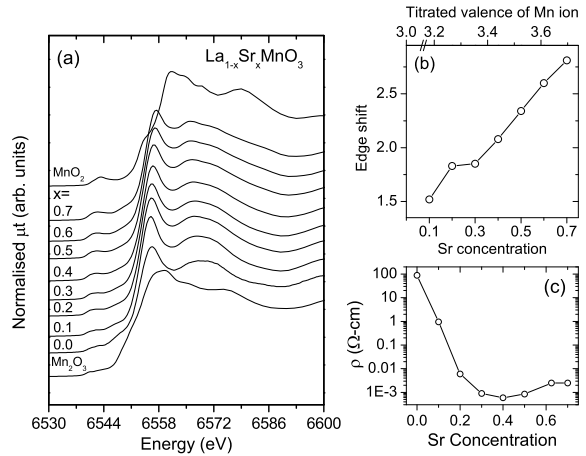}
\caption{(a)Experimental Mn K-edge x-ray absorption spectra for
Mn$_2$O$_3$, MnO$_2$ and La$_{1-x}$Sr$_x$MnO$_3$, $x$ = 0 - 0.7.
(b) Edge shift in Mn K-edge with respect to LaMnO$_3$ and (c)
resistivity at 300 K (from [4] and [23]) for
La$_{1-x}$Sr$_x$MnO$_3$.} \label{expt}
\end{figure}

The experimental XA spectra at Mn K-edge (6539 eV) for
La$_{1-x}$Sr$_x$MnO$_{3}$, $x$ = 0 - 0.7 together with the ones
for Mn$_2$O$_3$ (Mn$^{3+}$) and MnO$_2$ (Mn$^{4+}$) are shown in
figure \ref{expt}a. The edge-height for all the spectra is
normalized to unity. The edge energy E$_\circ$ has been taken as
the energy position of the first inflection point on the
absorption edge.  It is evident from the figure that E$_\circ$ of
all the La$_{1-x}$Sr$_x$MnO$_{3}$ spectra lie between the edge
energies for the absorption edge of Mn$_2$O$_3$  and MnO$_2$ with
Mn in 3+ and 4+ valence states, respectively.  All the absorption
spectra also exhibit three pre-edge features; these will be
discussed elsewhere.

The edge shift for La$_{1-x}$Sr$_x$MnO$_{3}$, $x$ = 0 - 0.7
measured from the absorption edge position of parent compound
LaMnO$_3$ is shown in figure~\ref{expt}b.  As seen from the figure
this edge shift increases monotonically with increase in Sr
concentration and the Mn valence as determined from the iodometric
titration. As remarked earlier, these shifts have both doping and
size contributions; below we discuss how these two contributions
can be separated using band structure calculations. In order to do
so, we first establish that the shift in the absorption edge
energy on Sr doping can be related with the shift in the CG of the
Mn 4$p$ band contributing to the absorption edge. Then we
establish that the total shift in CG is sum of shifts in CG due to
doping and size contributions explicitly for $x$ = 0.5 and 1.0, as
they are computationally less demanding. This additive character
of doping and size contributions to the edge shift is used to
obtain the doping contribution from the experimental edge shift
and calculated size contribution for all other compounds of the
series. Then we discuss how these contributions are used to
understand the transport behaviour of these compounds in a simple
manner.

\begin{figure}
\includegraphics[width=8.4cm]{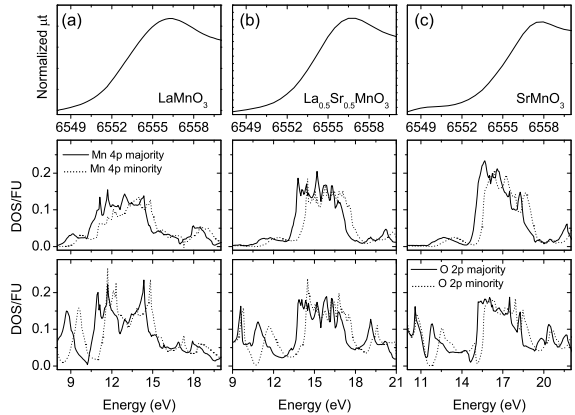}
\caption{Experimental Mn K-edge x-ray absorption edge (top panel),
calculated spin polarized partial density of states per formula
unit for Mn 4$p$ (middle panel) and O 2$p$ (lower panel) of (a)
LaMnO$_3$, (b) La$_{0.5}$Sr$_{0.5}$MnO$_3$ and (c) SrMnO$_3$.}
\label{band}
\end{figure}

In figure \ref{band}, the upper panels show the experimental Mn
K-edge absorption spectra for LaMnO$_3$,
La$_{0.5}$Sr$_{0.5}$MnO$_3$ and SrMnO$_3$, and the lower two sets
of panels show the calculated Mn 4$p$ and O 2$p$ partial density
of unoccupied states (PDOS) per formula unit for these compounds.
The calculated PDOS agree well with the reported ones by Ravindran
{\it et al.} \cite{ravi}. The PDOS of only $p$ symmetric states is
shown since for K-edge only $s \rightarrow p$ transition is dipole
allowed and thus is relevant for the present discussion. The
contribution of O 2$p$ unoccupied states is minimal to the
absorption edge as it involves a matrix element with wave
functions centred on spatially separated atoms; therefore, we
discuss only about the CG of Mn 4$p$ band. The edge position
corresponds to the difference between 1$s$ and 4$p$ levels; we
note that 1$s$ energy with respect to Fermi level hardly changes
($\sim$0.02 eV) with $x$, and the edge shift is given by that in
4$p$ band only. The value of average of CG of Mn 4$p$ majority and
minority bands for these three compounds is 12.99, 15.65 and 16.94
eV, respectively. Thus the shift in CG for
La$_{0.5}$Sr$_{0.5}$MnO$_3$ and SrMnO$_3$ with respect to
LaMnO$_3$ is 2.66 and 3.95 eV, respectively. Similarly, the Mn
K-edge energy for these compounds is 6551.88, 6554.22 and 6555.67
eV with corresponding shifts in E$_\circ$ for 50\% doped and
SrMnO$_3$ compounds being 2.34 and 3.79 eV, respectively.  The
shift in CG is thus approximately same (within $\sim$0.3 eV) as
the shift in absorption edge energy. Such behaviour is also seen
for shifts in average CG and E$_\circ$ for CaMnO$_3$ with respect
to LaMnO$_3$.  This indicates that the edge shift can be
understood in terms of the shift in the CG of Mn 4$p$ band
contributing to the main Mn K-edge and thus CG can be used to
monitor the changes in the edge shift. Such an identification of
the edge-shift with the shift in CG of the DOS of the contributing
band to the absorption edge would be applicable when this DOS has
'one peak' structure.  However, when the unoccupied DOS is spread
over different 'peaks' in the absorption edge region,
corresponding structures will be seen on the edge; a separate
inflection point can be identified for each structure which would
correspond to the CG of the appropriate energy domain of the band
structure.  This we have explicitly verified for binary compound
MnO whose Mn K-edge is structured.  Our calculations also show
that the total shift in average CG is less than 0.2 eV when
different magnetic ordering ({\it i.e.} FM, C-AFM, A-AFM) are
considered. This would indicate that edge shift does not depend
much on type of magnetic ordering in conformity with earlier
results \cite{ignatov}. Henceforth, correspondence of E$_\circ$
with CG obtained from the calculations for the FM phase only is
considered.

We now show using the instances of La$_{0.5}$Sr$_{0.5}$MnO$_3$ and
SrMnO$_3$ that the shift in CG can be used to separate out the
doping and size contributions to the edge shift. For other values
of $x$, it is rather involved as the necessary unit cell would be
too big and contain a much larger number of atoms. The band
structure calculations for $x$ = 0.5 compound were done using
following three lattice models to separate out the doping and size
contributions: (a) using Mn-O bond lengths of $x$=0.5 compound
obtained from EXAFS thus representing the size contribution with
respect to the parent compound LaMnO$_3$, (b) using the Mn-O bond
lengths of LaMnO$_3$ and replacing 50 \% of La atoms by Sr thus
representing the doping contribution with respect to LaMnO$_3$ and
(c) using bond lengths corresponding to
La$_{0.5}$Sr$_{0.5}$MnO$_3$ and replacing 50\% La atoms by Sr by
doubling the unit cell, thus considering both doping and size
contributions together. The shift in average CG of Mn 4$p$ band
with respect to that of LaMnO$_3$ obtained from (a) and (b) are
1.32 and 1.28 eV, respectively. The summation of these two shifts
is 2.60 eV,  approximately equal to a shift of 2.66 eV obtained
from (c). Similarly, the calculated values of doping and size
contributions for SrMnO$_3$ are 2.35 eV and 1.81 eV, respectively.
The summation of these two is 4.16 eV, which is close to the
calculated value of 3.95 eV when both doping and size
contributions are considered together. Thus the edge shift
$\Delta$ (and the shift in CG of Mn 4$p$ band) can be written as
$\Delta = E_d + E_s$ where $E_d$ and $E_s$ are doping and size
contributions to the edge shift, respectively. This indicates that
such a strategy could be adopted to determine these two
contributions to the edge shift. Therefore, we now calculate $E_s$
for rest of the compounds and estimate $E_d$ by subtracting it
from the edge shift observed experimentally.

\begin{figure}%[htbp]
\includegraphics[width=8.4cm]{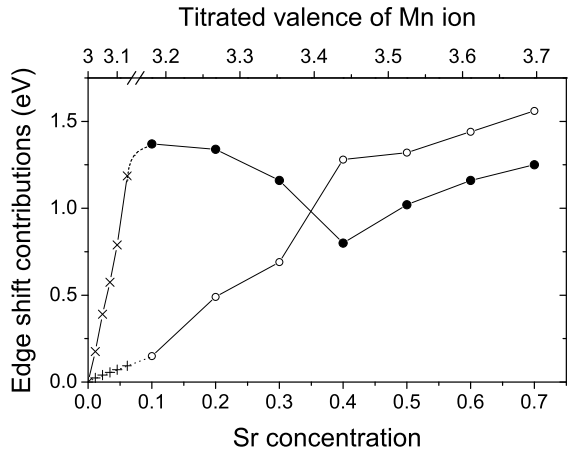}
\caption{The doping (close circles) and size (open circles)
contributions to the x-ray absorption Mn K-edge shifts for
La$_{1-x}$Sr$_x$MnO$_3$ with respect to LaMnO$_3$. The cross and
plus signs correspond to the expected behaviour of doping and size
contributions, respectively, when the valence state of Mn ions is
${<}\sim$ 3.15. } \label{contr}
\end{figure}

In figure \ref{contr}, the shifts in CG with respect to the CG of
LaMnO$_3$ due to size and doping contributions are shown by open
and solid circles, respectively, indicating that $E_s$ increases
monotonically with Sr concentration whereas $E_d$ first decreases
between $x$ = 0.1 - 0.4 and then increases as $x$ is increased to
0.7. On extrapolating $E_s$ for $x<$ 0.1, $E_s$ approaches zero as
$x$ tends to zero. This is shown by plus signs in the figure.  The
small rate of increase in $E_s$ in this region is consistent with
the observed very little change in Mn-O bond lengths between Mn
valence of 3.0 and $\sim$3.15 \cite{shibata}. The doping
contribution should also vanish at $x$ = 0 as the edge shift is
measured with respect to LaMnO$_3$.  This would imply that $E_d$
will increase in this region as indicated by the cross signs in
the figure.

A look at figure \ref{expt}c clearly shows that the behaviour of
resistivity with $x$ is very similar to the behaviour of $E_d$ for
$x \ge$ 0.1 - it first decreases with $x$ up to 0.4 and then
increases.  As discussed above the doping contribution $E_d$
arises due to changes in electronic configuration only keeping the
Mn-O bond lengths as in LaMnO$_3$. Thus doping contribution is
basically representing the effect of hole doping on edge shift. A
decrease in $E_d$ between $x$ = 0.1 - 0.4 would thus mean that on
doping there is a decrease in the number of localized holes.  This
result is surprising as one would have expected $E_d$ to increase
with $x$ as the number of holes increases with increase in Sr
doping.  It may be remarked here that all the doped holes  may not
be localized; they may exist simultaneously in localized and
delocalized states as suggested by Ramakrishnan {\em et al.}
\cite{tvr}.  We directly correlate $E_d$ with observed resistivity
by considering hole localization. The decrease in $E_d$ would mean
that holes delocalize between 0.1 $\le x \le$ 0.4.  Holes indeed
dominate the transport behaviour of manganites in this range of
doping as is evident from the Hall effect measurements
\cite{gordon,majumdar,asamitsu}. Since delocalization of holes
increases up to $x$ = 0.4, such a delocalization should result in
decrease in resistivity as has been observed experimentally
\cite{uru} (see figure \ref{expt}c). For $x >$ 0.4, the holes
start localizing at Mn sites as is evident from the increase in
$E_d$. This would imply that for higher compositions hole
localization shall lead to increase in resistivity. This is in
conformity with the observed increase in room temperature
resistivity \cite{hem}.

For $x <$ 0.1, $E_d$ should increase with $x$ as shown by the
extrapolation, indicating increased hole localization.  In the
first instance it would mean that resistivity should increase with
$x$ as observed for $x>$ 0.4. However, experimentally observed
behaviour of resistivity is opposite {\em i.e.} it decreases with
$x$.  This behaviour  can be understood by noting that the
compounds in this region are insulating and the conduction process
may be similar to that in semiconductors. Therefore, here the
effect of increase in number of holes due to doping may be the
dominating phenomenon contributing to the decrease in resistivity.

\section{Conclusions}

In conclusion, we have shown using a simple model based on first
principle calculations and experimental x-ray absorption spectra
that the energy shift in Mn K-edge on Sr doping in LaMnO$_3$ has
two contributions, namely doping and size contributions.  While
the doping contribution to Mn K- edge shift arises due to changed
number of valence electrons, the changes in Mn-O bond length leads
to size contribution.  It is observed that the observed doping
contribution with varying Sr concentration follows the observed
resistivity behaviour.  It is discussed that the changes in doping
contribution is directly related with the localization of charge
carriers thus bringing out the importance of hole localization in
understanding the electrical transport properties of these
compounds.

\acknowledgments RB and SKP thank UGC-DAE CSR for financial
support. AK thanks CSIR, Government of India for the senior
research associate position (pool scheme) during part of this
work.

%\newpage

\end{document}